\documentclass[epsf,aps,twocolumn,nofootinbib]{revtex4}
\usepackage{amssymb}

\usepackage{graphicx}
\usepackage{amsmath}
\usepackage{epsfig}

\def\B{\begin{eqnarray}}
\def\E{\end{eqnarray}}
\def\be{\begin{equation}}
\def\ee{\end{equation}}

\def\nn{\nonumber}
\def\x{\mathbf{x}}
\def\d{\mathrm{d}}
\def\yt{\tilde{y}}
\def\xt{\tilde{x}}

\def\mn{{\mu\nu}}
\def\dm{\partial_\mu}

\begin{document}

\title{Simple fixed-brane gauges in $S_1/\mathbb{Z}_2$ braneworlds}

\author{Samuel Webster{\footnote{e-mail:
      S.L.Webster@damtp.cam.ac.uk}}}

\affiliation{Department of Applied Mathematics and Theoretical
  Physics,\\
University of Cambridge, Wilberforce Road,\\
Cambridge CB3 0WA, UK}

DAMTP-2006-25

\begin{abstract}
For five-dimensional braneworlds with an $S_1/\mathbb{Z}_2$ orbifold
topology for the extra dimension $x^5$, we discuss the validity of
recent claims that a gauge exists where the two boundary branes lie at fixed
positions and the metric satisfies $g_{\mu 5}=\partial_5 g_{55}=0$
where $\mu$ labels the transverse dimensions. We
focus on models where the bulk is empty apart from a negative cosmological
constant, which, in the case of cosmological symmetry, implies the
existence of a static frame with Schwarzschild-AdS geometry. Considering the background case with the
branes moving apart after a collision, we show that such a
gauge can be constructed perturbatively, expanding in either the time
after collision or the brane velocity. Finally we examine how cosmological
perturbations can be accommodated in such a gauge.
\end{abstract}

\maketitle

\section{Introduction}
The large number of recent publications on braneworld cosmology is a testament
to the richness and difficulty of this comparatively new
field (see, for example, \cite{Brax:2004xh,Langlois:2002bb,Binetruy:1999ut,Binetruy:2001tc,Kim:2003pc,Maartens:2003tw}). 
Despite the large gap in complexity between possible String/M-Theory realisations and the greatly simplified models currently being
investigated, even the simplest models are extremely
rich in new physics. Many such five-dimensional toy models
\cite{Randall:1999ee,Lukas:1998tt,Lukas:1998qs,Lukas:1999yn,Shiromizu:2004ig,Kobayashi:2002pw,Brax:2000xk,Brax:2002nt}
are motivated by the
dimensional reduction of Ho\v{r}ava-Witten M-theory
\cite{Horava:1996ma} to five dimensions
but miss out most of the resulting large number of bulk degrees of freedom.The extra dimension is assumed to
have the topology of an $S_1/\mathbb{Z}_2$ orbifold, with two 3-branes
residing at the boundaries of the spacetime at the two fixed
points.

The increase in complexity from a
purely four-dimensional Universe to one confined to a brane embedded
in a five-dimensional bulk is considerable. We shall therefore mainly
focus on the Randall-Sundrum model \cite{Randall:1999ee}, where the bulk is
empty apart from a cosmological constant. We assume that the branes
contain no matter and carry equal and opposite tensions finely tuned with respect to the bulk cosmological constant such that their
induced cosmological constants vanish, although the conclusions of
this paper will not depend sensitively on these assumptions. In the
background case, where the transverse dimensions are assumed to
possess cosmological symmetry, the lack of non-gravitational degrees
of freedom in the bulk then implies the existence of a \emph{Birkhoff}
frame in which the bulk is manifestly static with
Anti-de-Sitter(AdS)-Schwarzschild geometry, parameterised by $C$
(related to the black hole mass) and the lengthscale $L$. The velocities of the branes
are then determined entirely by their positions, and the system can be solved
exactly even in the presence of matter on the branes; the only
deviation from four-dimensional Einstein gravity is the presence in
the Friedmann equation of quadratic terms in the brane stress-energy
and a dark radiation term proportional to $C$, which contains
all of the gravitational effects of the bulk.

When one comes to include cosmological perturbations \cite{vandeBruck:2000ju,Maartens:2004yc}, however, the
problem rapidly becomes intractable, largely due to the considerable
increase in number of the gravitational degrees of freedom in going
from four to five dimensions. To achieve any sort of analytic result
on the propagation of braneworld perturbations one needs to resort to
approximations. The simplest is to work to lowest order in the brane
velocities, assuming also that the energy content of the branes is
much less than their tensions. The result is a four-dimensional
effective scalar-tensor theory
\cite{Mendes:2000wu,Khoury:2002jq,Shiromizu:2002qr,Kanno:2002ia,Kanno:2002ia2,Brax:2002nt,Webster:2004sp,Palma:2004et},
valid at low energies, formulated for
example in terms of the induced
metric of one of the branes. The scalar degree of freedom, known as
the radion, corresponds to the distance between the two
branes. 

Attempting to go beyond the very limiting low-energy approximation
requires a method of solving the five-dimensional Einstein
equations. While a great deal of progress has been made recently
\cite{Palma:2004fh,deRham:2004yt,Calcagni:2003sg,McFadden:2005mq,Maartens:1999hf,Langlois:2000ns,Liddle:2001yq,Koyama:2004ap,deRham:2005qg,deRham:2005xv,Shiromizu:2002ve},
the problem is still very difficult. The five-dimensional analysis is
rendered simpler by choosing a gauge in which the branes are located
at fixed positions. The further assumptions that the coordinate $x^5$ parameterising the extra dimension is chosen such that
both the  cross terms $g_{5\mu}$ (where $\mu=0,1,2,3$ labels the
transverse time and space dimensions) vanish and $\partial_5
g_{55}=0$ result in a particularly useful gauge, henceforth referred
to as a \emph{simple fixed-brane} (SFB) gauge. Several publications \cite{Kanno:2002ia2,deRham:2005qg,deRham:2005xv,Shiromizu:2002ve}
have already assumed, without proof, the existence of SFB gauges to derive effective
theories with different, more useful regimes of validity than the low-energy
scalar-tensor theory. In particular they focused on the small-radion
limit pertinent to recent attempts to model the Big Bang as a brane
collision \cite{Khoury:2001wf,Khoury:2001bz,Gibbons:2005rt,Jones:2002cv,Turok:2004gb,Tolley:2003nx,Steinhardt:2002ih,Kanno:2002py}.

This paper is organised as follows. In \S\ref{5d formalism} we discuss
the general problem of constructing SFB gauges, specialising in
\S\ref{background} to the special case of background spacetimes with
cosmological symmetry. For the case of two branes moving apart after a
collision, initially considered in the Birkhoff frame, we construct in
\S\ref{t expansion} the required coordinate transformation as a power
series in the SFB time $t$, equivalent to an expansion in the
radion. We show that such a gauge can be found correct to arbitrary
orders in $t$, although there is in general no analytic expression for
the terms in the series and one must resort to numerical methods. We also comment on the convergence of this series and its
relation to the horizon structure of the bulk. A second method is
derived in \S\ref{C expansion}, namely an expansion in
$\sqrt{C}$, which is equivalent to the velocity of the branes. This
turns out to be simpler, with analytic expressions available at all
orders. Finally we discuss cosmological perturbations in
\S\ref{perturbations}, and show the existence of the gauge
transformations which will accommodate general perturbations into an
SFB gauge.

\section{General Formalism}
\label{5d formalism}
We take our spacetime to be an $S_1/\mathbb{Z}_2$ orbifold with two
boundary branes at the fixed points. We assume first of all the
existence of a Gaussian-Normal coordinate system relative to the
positive tension brane at $\yt=0$, with metric
\be
\label{gn metric}
\d s^2=\d \yt^2 + h_\mn\left(\xt,\yt\right)\d\xt^\mu \d\xt^\nu.\ee
Throughout this paper Greek indices will run over $0,1,2,3$ and denote
the transverse spatial $x^i$ and temporal $x^0$
coordinates, and $a,b,c$ etc. are five-dimensional indices
running over $0,1,2,3$ and $5$ for the orbifold dimension. The location of the positive-tension brane at $\yt_+=0$ is
not crucial for the discussion, of more importance is the form of the
metric with $g_{55}=1$ and $g_{5\mu}=0$ for the sake of simplicity in
the following argument.

Though the coordinate system (\ref{gn metric}) is particularly simple,
the second brane will, in general, lie on a spacetime-dependent
trajectory $\tilde{y}=\tilde{y}_-(x)$. Implementing the Israel junction conditions on
a moving brane is a complicated procedure \cite{Bucher:2004we}; far
simpler is to use a coordinate system in which the branes lie at fixed
positions, at the cost of losing the simple form of the metric
(\ref{gn metric}). A general coordinate transformation to bring the
branes to fixed positions will introduce off-diagonal
terms $g_{\mu 5}$ and a non-trivial $g_{55}$ to the metric. However, the
aim of this paper is to show the existence of a `simplest possible' SFB gauge where both
the non-diagonal metric terms vanish and $g_{55}$ has no dependence on
the extra dimension $y$
\be
\label{fixed gauge}
\d s^2=d(x)^2\d y^2+g_\mn(x,y)\d x^\mu \d x^\nu,\ee
in which the branes lie at the fixed positions $y=0$ and $y=1$. The
metric coefficient $d(x)$ can then be viewed as a four-dimensional
scalar field which plays the r\^ole of the radion, giving the proper
distance between the branes on curves of constant $x^\mu$. Note that, as
discussed in \cite{deRham:2005qg}, this definition of the radion is
frame dependent. As we shall see in \S\ref{background dynamics}, this
function is expected to be monotonically increasing (for the
homogeneous cosmological background) for the case of
two branes moving apart after a collision, and vanishes only at the
surface of collision.

Defining the diffeomorphism
\be
\label{f,g 5d}
\xt^\mu=f^\mu\left(x^\alpha,y\right),\
\yt=g\left(x^\alpha,y\right),\ee
we must then find $C^\infty$ functions $f^\mu$ and $g$ such that
\B
\label{diag}
\frac{\partial g}{\partial y}\dm
g+h_{\rho\sigma}\left(f^\alpha,g\right)\frac{\partial
f^\sigma}{\partial y}\dm f^\rho&=&0\\
\label{k}
\left(\frac{\partial g}{\partial
    y}\right)^2+h_\mn\left(f^\alpha,g\right)\frac{\partial
    f^\mu}{\partial y}\frac{\partial
    f^\nu}{\partial y}&=&d(x^\alpha)^2>0.\E
The new metric is then of the form (\ref{fixed gauge}). $g(x)$
    must then satisfy boundary conditions related to the positions
    $\yt_\pm(\xt)$ of the branes in the original frame (\ref{gn metric})
\be
\label{g bc}
g\left(x,0\right)=\tilde{y}_+\left(f(x,0)\right),\
g\left(x,1\right)=\tilde{y}_-\left(f(x,1)\right)\ee
in order to bring the positions of the branes in the new frame to $y=0$ for the
positive-tension brane and $y=1$ for the negative-tension brane. Note
that there is a residual gauge freedom of $x$-reparameterisation which
leaves both the form of the metric and the brane positions
invariant.

This system of non-linear, partial differential equations is extremely difficult to
tackle in general. However, since we are ultimately interested in cosmological
braneworlds, we will at first only need to consider these equations for the
homogeneous case, effectively reducing the dimensionality of the problem from five to two. Furthermore, this paper is only concerned with
Randall-Sundrum style braneworlds where the bulk contains only a
(negative) cosmological constant; staticity of the bulk then follows in the
homogeneous case from Birkhoff's theorem.

This static \emph{Birkhoff Frame} is of the form (\ref{gn metric})
and we investigate its properties in the next section. We will
describe two methods of constructing the coordinate transformation
satisfying (\ref{diag}) and (\ref{k}) for the background, and then turn our
attention to cosmological perturbations.

\section{Background Formalism}
\label{background}
\subsection{Birkhoff Frame Dynamics}
\label{background dynamics}
In this section we re-derive some standard results on the propagation
of branes in the static Birkhoff frame background \cite{deRham:2005xv,deRham:2005qg,Binetruy:1999hy}, with metric
\be
\label{birkhoff}
\d s^2=\d Y^2-n^2(Y)\d T^2+a^2(Y)\d\x^2 \ee
where
\B
\label{a}
a^2(Y)&=&e^{-2 Y/L}+\frac{C}{4}\ e^{2 Y/L}\\
\label{n}
n^2(Y)&=&a^2-\frac{C}{a^2}\Rightarrow a'(Y)=-n(Y)/L
\E
The bulk geometry is Schwarzschild-AdS, with the AdS lengthscale $L$
related to the cosmological constant $\Lambda$ by $\Lambda=-\frac{6}{L^2}$ and the parameter $C$ proportional to the
mass of the bulk black hole. We assume that the brane tensions are
fine-tuned to the usual values $\lambda_\pm=\pm\frac{6}{L}$
where the five-dimensional gravitational constant has been set to
unity for convenience. The spatial curvature is assumed to vanish for
simplicity.

In this frame the branes will not be static, but follow trajectories
$Y_\pm(T)$. Assuming for the moment that the branes are empty, the
Israel junction conditions are simply
\be
\label{junction}
K^{a \pm}_b=-\frac{1}{L}\delta^a_b\ee
where $K^{a\pm}_b$ are the extrinsic curvatures of the branes. The
space-space components of this equation then give the first-order equations of motion
\cite{deRham:2005xv,deRham:2005qg}
\be
\label{dYdT}
\left(\frac{\d Y_\pm}{\d
  T}\right)^2=C\frac{n_\pm^2}{a_\pm^4}\ee
where $a_\pm=a(Y_\pm)$ etc. Since we are interested in brane
  collisions we assume that $\d Y_\pm/\d T$ are of opposite signs, specifically
  that $\d Y_+/\d T <0$ and $\d Y_-/\d T>0$.  The proper time $t_\pm$ on the branes
  can then be determined, giving the world velocities and accelerations as
\B
\label{world velocity}
\frac{\d Y_\pm}{\d t_\pm}=\frac{\sqrt{C}}{a_\pm n_\pm}&,&\quad\frac{\d T_\pm}{\d t_\pm}=\frac{a_\pm}{n_\pm^2}\\
\label{acceleration}
\frac{\mathrm{D}^2 Y_\pm}{\mathrm{D} t_\pm^2}=-\frac{a_\pm}{n_\pm
  L}&,&\quad \frac{\mathrm{D}^2 T_\pm}{\mathrm{D}
  t_\pm^2}=-\frac{\sqrt{C}}{n^2_\pm a_\pm L}\E
The above equations imply that, for empty,
  fine-tuned, spatially-flat branes, one cannot achieve a consistent
  embedding into the negative mass Schwarzschild-AdS spacetime with
  $C<0$. It also implies that, under the same conditions, the $C=0$
  solution is the trivial case where the branes remain static and the frame reduces
  to the usual AdS warp-factor profile, although from
  (\ref{acceleration}) we see that the branes are still accelerating.

 We shall therefore henceforth only consider
  the case of \emph{positive} spacetime mass $C>0$. In this case there is a horizon in the bulk where $n(Y)$ vanishes, given by
\be
\label{horizon}
Y_\mathrm{hor}=\frac{L}{2}\log\frac{2}{\sqrt{C}}.\ee
The coordinate system (\ref{birkhoff}) will only be valid in a
particular coordinate patch bounded by the horizon. One would
therefore expect any results in
this paper derived from this coordinate system only to be valid
up to the point where one of the branes hits the horizon (in other
words, for the physical slice of spacetime bounded by the two branes
to develop a horizon).
Since $T$-translation is a symmetry of the
metric (\ref{birkhoff}) we can assume that the brane collision occurs
at $Y=Y_0$, $T=0$. Fig.~1 shows the branes' trajectories in the
Birkhoff frame with $L=C=1,Y_0=0$. Note that a brane crossing the horizon will
do so in finite proper time but will require an infinite $\Delta T$,
since from (\ref{dYdT}) we see that the brane velocity with respect to the
Birkhoff time $T$ tends to zero as the
horizon is approached. 
The corresponding picture as a Penrose diagram is illustrated in Fig.~2, showing the
region of the extended Schwarzschild-AdS spacetime
corresponding to the coordinate patch (\ref{birkhoff}). In this
picture, the horizon at $Y=Y_\mathrm{hor}$ can be interpreted as the event
horizon of the bulk black hole. Note that the physical spacetime is
just the slice bounded by the two branes (or, rather, two copies of it
due to the orbifold nature of the complete spacetime).

Importantly, it follows from (\ref{dYdT}) and (\ref{world velocity}) that the
Friedmann equation on the brane is simply $H^2_\pm=C/(L^2 a_\pm^4)$,
i.e. the background geometry is finite at the collision \cite{Kanno:2002py}. The event is,
however, singular from a five-dimensional perspective, in that the
fifth dimension momentarily vanishes. In general perturbations will
diverge due to the blue-shifting of their energies as the radius of
the fifth dimension vanishes, although several authors recently have
claimed that this need not be an obstacle to the propagation either of cosmological
perturbations \cite{Tolley:2003nx} or, more fundamentally, string
excitations \cite{Turok:2004gb,Niz:2006ef} across the bounce.

\begin{figure}[!h]
\begin{center}
\includegraphics[width=9cm]{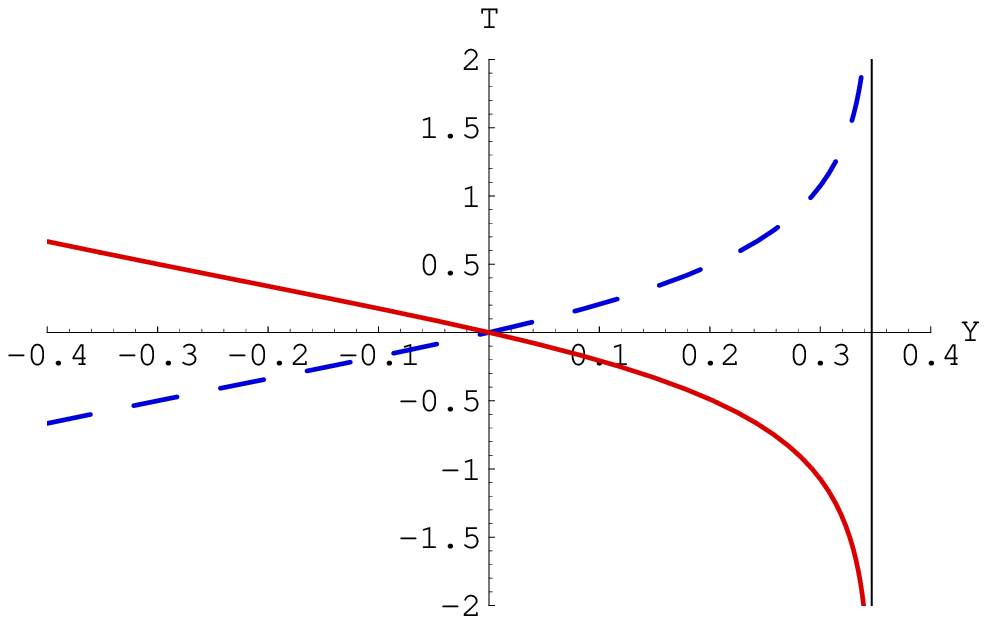}
\caption{Background brane trajectories for $L=C=1$,$Y_0=0$. The
  trajectory of the positive- \mbox{(negative-)} tension brane is the
  solid (dashed) line, and the event
  horizon at $Y=Y_\mathrm{hor}$ is denoted by the solid vertical line.}
\vspace{5mm}
\includegraphics[width=8cm]{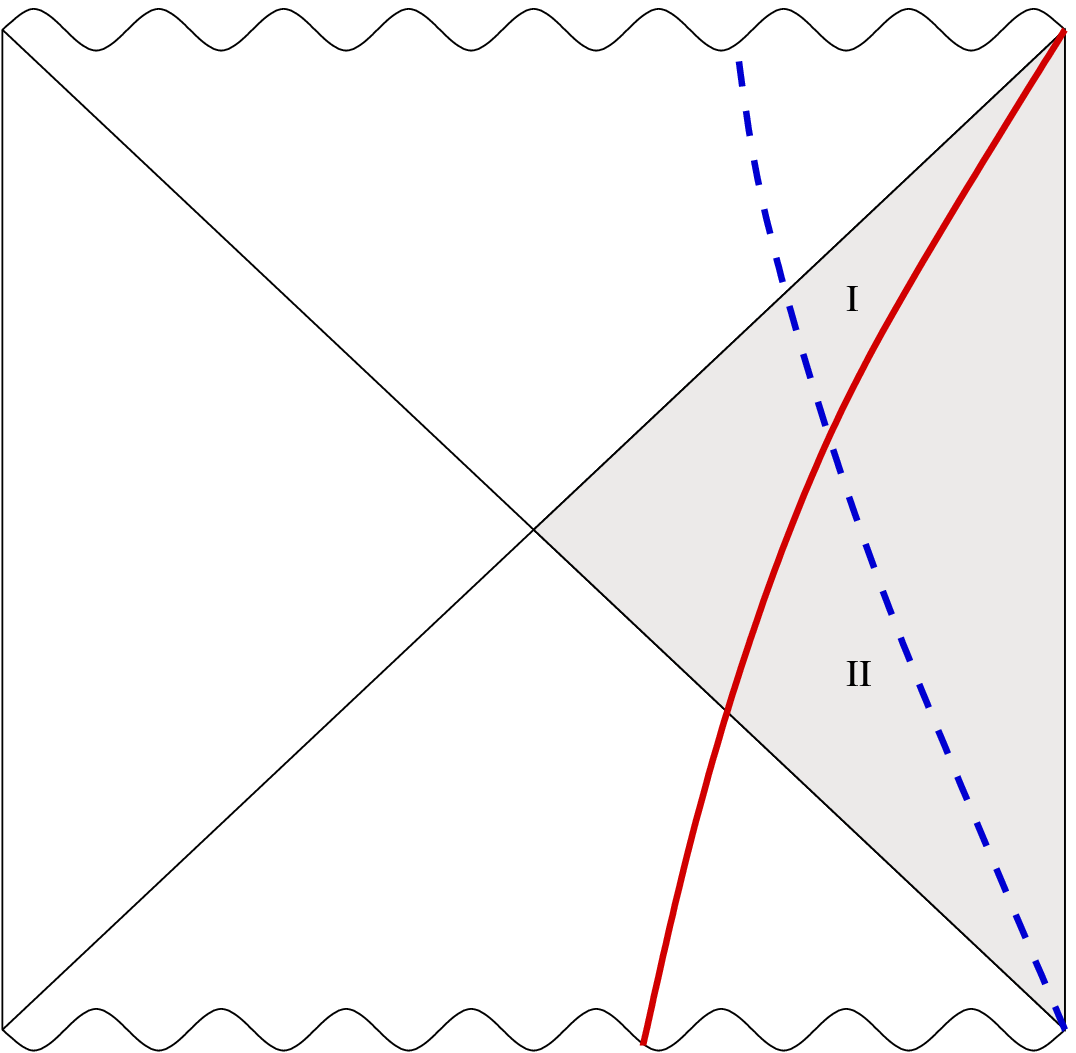}
\caption{The Penrose diagram corresponding to Fig.~1, i.e. for the case $C>0$. The shaded area denotes the
  region of extended Schwarzschild-AdS covered by the
  coordinate system (\ref{birkhoff}). The trajectory of the positive- \mbox{(negative-~)}
  tension brane is denoted by the solid (dashed) line. The physical
  region of spacetime is that bounded by the two branes. The diagonal lines bounding
  the coordinate patch represent the horizon $Y=Y_\mathrm{hor}$. Region I therefore represents the portion
  of the future of the collision which the $Y,T$ coordinates cover.}
\end{center}
\end{figure}

\subsection{Gauge Transformation}
We now simplify the analysis of \S\ref{5d formalism} to the
homogeneous, static background of \S\ref{background dynamics}.
The metric components (\ref{birkhoff}) and the brane trajectories
(\ref{dYdT}) do not, by homogeneity, depend on the spatial coordinates $x^i$.
We can therefore just leave these coordinates alone,
\mbox{$f^i\left(x,y\right)=x^i$}, and focus on the
two-dimensional $Y,T$ portion of the metric. As before, we consider a diffeomorphism
\be
\label{fg 2d}
T=f\left(y,t\right),\ Y=g\left(y,t\right).\ee
The conditions (\ref{diag}) and (\ref{k}) on the form of the new
metric become
\B
\label{eqn 1}
g_y g_t&=&n(g)^2 f_y f_t\\
\label{eqn 2}
g_y^2&=&n(g)^2 f_y^2+d(t)^2,\E
giving
\be
\label{fixed metric}
\d s^2=d(t)^2\d y^2 - b\left(y,t\right)^2\d t^2 + a\left(y,t\right)^2
\d \x^2\ee
where
\[b\left(y,t\right)^2=n(g)^2f_t^2-g_t^2.\]
It is at this point that the assumption
that the metric (\ref{birkhoff}) is static (i.e. that we are concerned
only with Randall-Sundrum-style braneworlds, with no non-gravitational
degrees of freedom in the bulk) comes in. This implies that $n$ is
only a function of $g$, rather than $n(f,g)$ as would be expected more
generally. One can then eliminate $f$ entirely using (\ref{eqn 1}) and (\ref{eqn
  2}), giving (up to an arbitrary sign)
\B
\label{fy}
f_y&=&\frac{1}{n(g)}\sqrt{g_y^2-d(t)^2}\\
\label{ft}
f_t&=&\frac{1}{n(g)}\frac{g_y g_t}{\sqrt{g_y^2-d(t)^2}}\E
The Jacobian of this coordinate transformation must not vanish, which
implies from (\ref{fy}) and (\ref{ft}) that
\[d(t)^2 g_t\neq0.\] We can therefore conclude that $g_t\neq 0 $ away
from the collision which, by convention, we shall assume occurs at $T=0$. In
the $(y,t)$ plane, the collision corresponds to the vanishing of the
metric component $d(t)$, and also to the curve $f(y,t)=0$. This curve has
zero proper length, lying at constant $t$, and wlog we can take the
collision to correspond to the line $t=0$ on which $f$ vanishes and
$g$ takes the constant value $Y_0$.\\
As we saw in \S\ref{background dynamics}, the distance between the
two branes is a monotonically-increasing function after the
collision. This implies that one can exploit the
$t$-reparameterisation of the metric to fix
\be
\label{t gauge fix}
d(t)=t.\ee
It turns out that this gauge choice ensures the existence of a
perturbative expansion of the functions $f$ and $g$ in powers of
$t$, as well as corresponding to the linear approximation for close
branes used in \cite{deRham:2005xv}.\\
Continuity implies that $f_{yt}=f_{ty}$, which from (\ref{fy}) and
(\ref{ft}) reduces to a second-order partial differential equation for $g\left(y,t\right)$:
\be
\label{g''}
g_t g_{yy}+t-\frac{g_y^2}{t}+\frac{n'(g)}{n(g)}\left(g_tg_y^2-t^2 g_t\right)=0\ee
Writing (\ref{dYdT}) as
\[
\frac{\d Y_\pm}{\d T}=\mp v\left(Y_\pm(T)\right),\]
we find
\B
\label{gt0}
\left.g_t\right|_{y=0}&=&\frac{\partial}{\partial
  t}Y_+\left(f(0,t\right)=-\left.v(g)f_t\right|_{y=0}\\
\label{gt1}
\left.g_t\right|_{y=1}&=&\frac{\partial}{\partial
  t}Y_-\left(f(1,t\right)=+\left.v(g)f_t\right|_{y=1}\E
Substituting (\ref{ft}) for $f_t$ we find, for example,
\B
\left.g_t\right|_{y=0}&=&-\left.\frac{v(g)}{n(g)}\frac{g_y\,g_t}{\sqrt{g_y^2-t^2}}\right|_{y=0}\nn\\
\Rightarrow \left.g_y^2\right|_{y=0}&=&\left.\frac{ n(g)^2 t^2
}{n(g)^2-v(g)^2}\right|_{y=0}\nn\E
Using (\ref{n}) and (\ref{dYdT}) then gives the boundary conditions
\be
\label{gy bc}
\left. g_y\right|_{y=0,1}=t\left.\frac{a(g)}{n(g)}\right|_{y=0,1}\ee
together with
\[g\left(y,t=0\right)=Y_0.\]
Note that the sign of $g_y$ in (\ref{gy bc}) is specified by the
choice $f_t>0$, $g_y>0$, i.e. that the orientation of the new coordinate
system is the same as the old one. The fact that the branes are moving
in opposite directions, not immediately apparent from the boundary
conditions since the $-$ sign in (\ref{gt0}) has been lost, follows
from the choice of gauge (\ref{t gauge fix}).

\section{Perturbative expansion in $t$}
\label{t expansion}
We now proceed to solve (\ref{g''}) subject to the boundary conditions
(\ref{gy bc}) as a perturbative expansion in powers of
$t$. Specifically, we write
\be
\label{power}
g\left(y,t\right)=Y_0+\sum_{n=1}^\infty g_n(y) t^n\ee
where the boundary conditions (\ref{gy bc}) will specify the
derivatives of $g_n$ at $y=0,1$ in terms of the values there of the lower-order
$g_i$ $(i<n)$. (\ref{g''}) becomes a set of ordinary differential
equations for the $g_n$; the aim therefore is to identify the correct
initial value $g_n(0)$ for which the two boundary conditions
are simultaneously satisfied.
Once $g(y,t)$ has been determined to the required order, $f(y,t)$ can
in principle be recovered from (\ref{ft}) with boundary condition $f(y,0)=0$.
\subsection{First order}
The $O(t)$ contribution to (\ref{g''}) gives
\be
\label{g1''}
g_1 g''_1-{g'_1}^2+1=0\ee
with (\ref{gy bc}) implying
\be
\label{g1bc}
g'_1(0)=g'_1(1)=\frac{a_0}{n_0}\ee
where $a_0=a(Y_0)$ etc. We then obtain the result
\be
\label{g1}
g_1(y)=\frac{1}{2\alpha}\sinh\left(\alpha\left(2y-1\right)\right),\quad
\cosh\alpha=\frac{a_0}{n_0}\ee
The corresponding result for $f$ is
\be
f_1(y)=\frac{1}{2\alpha
  n_0}\cosh\left(\alpha\left(2y-1\right)\right).\ee
This is just the relevant coordinate transformation in the approximation of
constant brane velocities, as detailed in \cite{deRham:2005xv}.
\subsection{Second order}
Before examining the existence of $g_n$ for arbitrary $n$, we consider
the simpler case of $n=2$.
\be
\label{g2''}
g_1 g''_2-2g'_1 g'_2+2g_2 g''_1+\frac{n'(Y_0)}{n(Y_0)}\left(g_1 g'^2_1-g_1\right)=0\ee
Expanding out
\be
\label{boundary series}
K\left(g(y,t)\right)=K(Y_0)+t\,g_1(y)K'(Y_0)+O(t^2)\ee
for an arbitrary analytic function $K$, we identify the boundary conditions as
\[
g'_2(y=0,1)=\left.\left(\frac{a}{n}\right)'\right|_{Y=Y_0} g_1(y=0,1)\]
giving
\be
\label{g2bc}
g'_2(1)=-g'_2(0)=\frac{C \sinh\alpha}{L\alpha n_0^2 a_0^2}.\ee
There is no simple analytic general solution to this equation, so one
must proceed numerically. The task is to find the value of $g_2(0)$
such that, after integrating (\ref{g2''}) subject to the initial
conditions at $y=0$, $g'_2(1)$ is given by (\ref{g2bc}), for example
using a bisection procedure. The existence of $g_2$ is discussed
below.

\subsection{General $n$}
For tuned tensions, i.e. the boundary conditions (\ref{gy bc}), the
$g_n$ all have definite parity $(-1)^n$ about $y=1/2$. To show this,
we substitute the power series (\ref{power}) into (\ref{g''}) and
consider the coefficient of $t^n$, giving
\be
\label{gn''}
g_1 g''_n - 2 g'_1 g'_n+ n g''_1 g_n = P_n(y)\ee
where $P_n(y)$ is composed of a finite sum of products of the $g_i$
and their first or second derivatives, but only containing $g_i$ for
$i<n$ (c.f. (\ref{g2''})). Examining the contributions from the various terms in
(\ref{g''}) individually, we write symbolically (i.e. ignoring
numerical constants)

\[
g_t g_{yy}\sim\sum_{k=0}^\infty t^k\left( \sum_{m=1}^k g''_m\,
g_{k+1-m}\right)\supset\sum_{m=2}^{n-1}g''_m\, g_{n+1-m}\]
where the symbol $\supset$ denotes the terms in the coefficient of $t^n$ which do
not contain $g_n$, i.e. the contribution to the RHS of (\ref{gn''}).
Working under the inductive hypothesis that $g_m$ has parity $(-1)^m$
for $m<n$ (which is obviously true for $n=0$ and $n=1$), we see that
each summand has the parity $(-1)^{n+1}$. Similarly, $-\frac{1}{t^2} g_y^2$ makes a contribution
\[
\frac{1}{t} g_y^2\supset \sum_{m=2}^{n-1} g'_m\,g'_{n+1-m}\]
Again, the parity of these terms is $(-1)^{n+1}$ by
the inductive hypothesis. Next, we examine $g_t g_y^2-t^2 g_t$,
which is multiplied by a function of $g$ in (\ref{g''}), so we will
need all orders up to $n$ in its expansion. We find that
\B
g_t g_y^2&\sim&\sum_{k=2}^\infty t^k \left( \sum_{m=1}^{k-1}
\sum_{p=1}^{k-m} g_m\, g'_p\, g'_{k+1-m-p}   \right)\nn\\
\label{term 3}
&\sim& \sum_{k=2}^{n} t^k H_k(y)+O(t^{n+1})\nn\E
where $H_k(y)$ is a function of parity $(-1)^{k+1}$ involving only $g_m$ for $m<k$; there is
no contribution at order $t^n$ or below from $g_n$. $t^2 g_t$ is
obviously also of this form, so we just need to look at the
$n'(g)/n(g)$ term. Any function $K(g)$ with a power series expansion can be
written symbolically in the form
\be
\label{function expansion}
K(g)\sim K(Y_0)+\sum_{k=1}^\infty t^k\left(\sum_{m=1}^k\  \sum_{i_j>0,\sum
  i_j=k}g_{i_1}...g_{i_k}\right),\ee
which can be re-expressed as
\[
K(g)\sim\left(\sum_{k=0}^n t^k G_k(y)\right) + g_n(y) t^n + O(t^{n+1})\]
where $G_k(y)$ is a sum of products of $g_i$ with $i<k$, with total
parity $(-1)^k$. Putting this all together, we see that the RHS of
(\ref{gn''}) can be written as
\B
P_n&\sim& \sum_{m=2}^{n-1}g''_m\, g_{n+1-m} + \sum_{m=2}^{n-1}
g'_m\,g'_{n+1-m}\nn\\
&&+\sum_{m=0}^{n-2}G_m(y) H_{n-m}(y).\E
Each summand in each term has the same parity,
$(-1)^{n+1}$. Therefore, looking back to (\ref{gn''}), we see that the
whole equation is invariant under parity if $g_n$ has parity
$(-1)^n$. Whether or not this is the case depends on the boundary
conditions; in the fine-tuned case, the boundary conditions are just
that
\[
g'_n(y_*)=F_{n-1}(y_*)\]
where $y_*=0,1$ and
\[
\frac{a\left(g(y,t)\right)}{n\left(g(y,t)\right)}=\sum_{m=0}^\infty
  F_n(y) t^n\]
Note that this equation implies that $g'_n$ at the boundaries is
  specified entirely in terms of the values there of the $g_m$ for $m<n$.
From (\ref{function expansion}), $F_{n-1}(y)$ has parity
  $(-1)^{n-1}$. Hence
\be
\label{boundary parity}
g'_n(1)=(-1)^{n-1} g'_n(0)\ee
and the inductive hypothesis is proven.
One can now show the existence of $g_n$ for $n$ odd - since the
function will be odd, it and its second derivative must vanish at
$y=1/2$. The equation of motion will then prescribe the value of
$g'_n(1/2)$ which, together with $g_n(1/2)=0$, will give a unique solution.

For even $n$, one must use a more general argument. We are searching
for the value of $g_n(0)$ which gives a particular value for
$g'_n(1)$. Taking $g_n(0)=N$ for $N$ very large, but $g'_n(0)$ to its
finite, prescribed value,  the LHS of (\ref{gn''})
will dominate over the RHS apart from at the finite number of points
where either $g_n$ or its derivatives vanish. Therefore,
\be
\label{gn asymptotic}
g_n(y)\rightarrow N h_n(y)\ \mathrm{as}\ N=g_n(0)\rightarrow
+\infty\ee
where $h_n$ is the solution to the initial value problem
\[
g_1 f''_n-2 g'_1 f'_n+n g''_1 f_n=0,\quad f_n(0)=1,\ f'_n(0)=0\]
Provided that $h'_n(1)\neq 0$ (which can be shown numerically for
specific values of $n$), we then have that
$g'_n(1)\rightarrow\pm\infty$ as $g_n(0)\rightarrow+\infty$,
$g'_n(1)\rightarrow\mp\infty$ as $g_n(0)\rightarrow-\infty$. Hence, by
continuity, there will be (at least one) intermediate value of
$g_n(0)$ where $g'_n(1)$ takes the required value. Hence we can always
find a $g_n$ satisfying both (\ref{gn''}) and the two
boundary conditions for $g'_n$ at $y=0,1$ following from (\ref{gy
  bc}).

The above argument is not sensitive to the precise nature of the
boundary conditions; in other words, in the case of general brane
tensions and brane matter content, the boundary conditions (\ref{gy
  bc}) would be modified, but one would still expect to be able to
find suitable functions $g_n(y)$ to perform the coordinate transformation
to the SFB frame.

 However, the entire analysis is dependent
on the existence of a static bulk frame such as (\ref{birkhoff})
(though the precise nature of the functions $a$ and $n$ will, again,
not affect the argument). It is not clear if this result could be
generalised to more complicated braneworlds where the bulk is not
empty, such as dilatonic or supergravity-inspired models with bulk
scalar fields \cite{Shiromizu:2004ig,Kobayashi:2002pw,Brax:2000xk,Brax:2002nt,Mennim:2000wv}.

\begin{figure}[!h]
\begin{center}
\includegraphics[width=9cm]{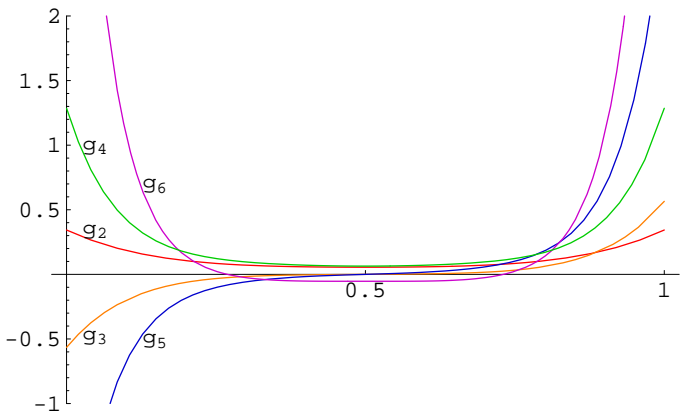}
\caption{$g_2$ to $g_6$ plotted for $L=1,C=1,Y_0=0$. Note the
  alternating parities.}
\vspace{5mm}
\end{center}
\end{figure}

\begin{figure}[!h]
\begin{center}
\includegraphics[width=9cm]{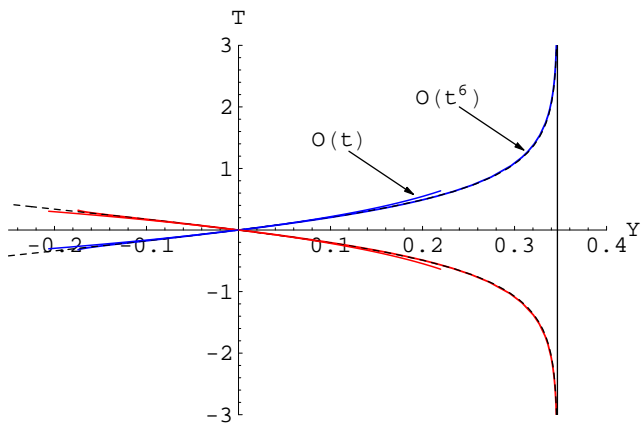}
\caption{Comparison of the brane trajectories computed using the coordinate transformation function $g(y,t)$ up to
  $O(t^6)$ with the exact result. The dotted curves correspond to the
  exact trajectories as in Fig.~1, showing highly accurate agreement
  with the numerical results. For comparison the corresponding
  trajectories computed working only to first order in $t$ are also
  shown for the same range of $t$, with clear differences.}

\vspace{5mm}
\end{center}
\end{figure}

\subsection{Numerical results and convergence}
The horizon structure of the bulk has
ramifications for the regions of validity of the coordinate system and
the radius of convergence of the power series (\ref{power}). For $C>0$
there is a horizon, whilst for $C<0$ there is a naked singularity and
no consistent embedding of flat, empty, finely-tuned branes. The
$C=0$ case, for fine-tuned branes, represents the trivial evolution
where the branes remain fixed at their original positions, and the
Birkhoff frame automatically satisfies the required gauge conditions
(\ref{fixed gauge}). We therefore only need consider the case $C>0$.

The first six non-trivial $g_n$ have been computed numerically using a bisection procedure to search
for $g_n(0)$ in the specific case of $Y_0=0$, $L=C=1$. The resulting
$g_2$ to $g_6$ are plotted in Fig.~3.

Fig.~4 gives the calculated brane trajectories
back in the $(Y,T)$ frame; for example, $g(0,t)$ gives $Y_-(t)$, and
integrating (\ref{ft}) with $d(t)=t$ gives $f(0,t)=T_-(t)$, hence
one can plot the trajectories $Y_\pm(t)$ and compare them with the
exact solutions. The accuracy of the agreement is high, as can be seen
from the Fig.~5. Also plotted are the trajectories calculated using only the first-order expression for $g(y,t)$, with the same range of $t$; these clearly differ wildly from the higher-order results as the horizon is approached.

Depending on the initial conditions, either the positive- or negative-tension brane will hit this horizon at a finite value of $t=t_h$. For the initial conditions chosen above, this corresponds to $t_h\approx 0.36$. At this point, $n(g)$ will vanish at the relevant boundary, and so (\ref{g''}) will break down and one would not expect the coordinate system (\ref{fixed metric}) to be valid beyond this value of $t$. In particular, one would expect this value of $t$ to correspond to the radius of convergence of the series (\ref{power}). It is clear from Fig.~3 that the boundary values of $g_n(y)$ are increasing in modulus as $n$ increases, which is a manifestation of the lack of convergence of the series for large $t$ in this case.

In the fine-tuned, empty-brane case with boundary conditions (\ref{gy bc}), the symmetry of the problem (i.e. the manifest reflection symmetry under $Y\rightarrow-Y$ in the trajectory plots) implies that the same coordinate system will also be valid before the collision, for $-t_h>t>0$.

The advantage of an expansion in the SFB time $t$ is that it is
ideally adapted to working in the limit of close branes, a regime of
considerable interest recently in the literature
\cite{McFadden:2005mq,deRham:2005qg,deRham:2005xv,Shiromizu:2002ve,Battye:2003ks,Leeper:2005pg}.
In effect, an expansion in $t$ is equivalent to an expansion in the
distance between the two branes. Working even just to first order in
$t$ or $d$ whilst making no other approximations, one can obtain
interesting, non-perturbative results unattainable using the moduli space approximation
\cite{deRham:2005qg,deRham:2005xv}. Whether or not the full series
(\ref{power}) converges is then not important if one simply wants to
proceed perturbatively in $t$.

\section{Perturbative expansion in C}
\label{C expansion}
An alternative method of finding the required gauge (\ref{fixed
  gauge}) is to consider an expansion in the velocity of the
  branes (\ref{dYdT}). Specifically, we define the parameter
\be
\label{h}
h=\sqrt{C}\ee
and expand in this rather than in $t$. This amounts to an expansion
about the $C=0$ AdS spacetime where the (finely-tuned) branes are at
fixed positions in the Birkhoff frame (\ref{birkhoff}). This method
turns out to be conceptually and analytically simpler than the
expansion in $t$, though it is of less direct use for the near-brane
regime, being an expansion about a static configuration. In this sense, working to linear order in $h$ will
reproduce the results of the moduli space approximation.

As discussed below, we will in fact expand about the state where the branes are
static and coincident, in order to simplify the analysis somewhat and
to tie in with the rest of the paper assuming that the branes are
moving apart from an initial collision. It would not, however, be
difficult to generalise the results to a different initial configuration.

\subsection{Formalism}
Firstly we define a new coordinate $Z=e^{-Y/L}$, in
terms of which the Birkhoff frame metric can be written
\B
\label{h birkhoff}
\d s^2&=&\frac{L^2}{Z^2}\d Z^2 - n(Z)^2\d T^2 + a(Z)^2 \d\x^2\\
a(Z)^2&=&Z^2+\frac{h^2}{4 Z^2}\nn\\
n(Z)^2&=&a(Z)^2-\frac{h^2}{a(Z)^2}\E
For $h=0$ the branes are static, at $Z=Z^{(+)}_0$ and
$Z=Z^{(-)}_0$. Since we are interested in modelling the aftermath of a
collision we shall take $Z^{(\pm)}_0=Z_0$. Whilst simplifying the
resulting analysis considerably one must then remember that, for
$h=0$, the metric will be singular. Therefore we shall expect $g_{yy}$ in the SFB frame
(\ref{fixed metric}) to vanish for $h=0$.
Bearing this in mind, we then perform a coordinate transformation as
before, keeping the $t$-gauge general for the time being (as opposed
to the choice $d(t)=t$ of the previous section). Expanding in $h$, we
write
\B
\label{Z}
Z(y,t)&=&Z_0\left(1+\sum_{n=1}^\infty h^n Z_n(y,t)\right)\\
\label{T}
T(y,t)&=&t+\sum_{n=1}^\infty h^n T_n(y,t),\E
where for $h=0$ we can set $T=t$ since the branes are already at fixed
positions. We then solve
\[
\partial_y g_{yy}=g_{yt}=0\]
order by order to determine the functions $Z_n(y,t)$ and $T_n(y,t)$,
together with the boundary conditions (\ref{dYdT}),
\be
\label{Z bc}
\frac{\d Z_\pm}{\d T}=\pm Z_\pm\frac{h}{L}\frac{n(Z_\pm)}{a(Z_\pm)^2}\ee

\subsection{First order}
In the $(y,t)$ frame, the metric becomes
\B
\label{yt metric}
\d s^2&=&\left(\frac{L^2}{Z^2}Z_y^2-n(Z)^2T_y^2\right)\d y^2\\
&&+2\left(\frac{L^2}{Z^2}Z_y Z_t-n(Z)^2T_y T_t\right)\d y\d t\nn\\
&&+\left(\frac{L^2}{Z^2}Z_t^2-n(Z)^2T_t^2\right)\d t^2+a(Z)^2\d\x^2\nn\E
and so, using (\ref{Z}) and (\ref{T}), we obtain
\B
\label{dygyy1}
\partial_y g_{yy}&=&2\left(L^2 \dot{Z}_1
Z''_1-Z_0^2\, \dot{T}_1 T''_1\right) h^2+O\left(h^3\right)\\
\label{gyt1}
g_{yt}&=&-Z_0^2 \dot{T}_1h+O(h^2).\E
Next we examine the boundary conditions (\ref{Z bc}),
\B
\frac{\d Z_+}{\d T}&=&\frac{\partial_t Z(0,t)}{\partial_t T(0,t)}=Z_0
\dot{Z}_1(0,t) h+O(h^2)\nn\\
&=&Z_+\frac{h}{L}\frac{n(Z_+)}{a(Z_+)^2}=\frac{h}{L}+O(h^2)\nn\E
giving
\be
\label{z1bc1}
\vspace{-3mm}
\dot{Z}_1(0,t)=\frac{1}{L Z_0}\ee
and, similarly,
\be
\label{z1bc2}
\dot{Z}_1(1,t)=-\frac{1}{L Z_0}.\ee
(\ref{gyt1}) gives the simple condition that $\dot{T_1}=0$ and,
choosing the collision to occur at $t=0$, implies
\be
\label{t1}
T_1(y,t)=0.\ee
The condition (\ref{dygyy1}) on $g_{yy}$ reduces to
\[
Z''_1=0\]
which, together with the boundary conditions (\ref{z1bc1}),
(\ref{z1bc2}) and $Z_1(y,0)=Z_0$, gives
\be
\label{z1}
Z_1(y,t)=\frac{t}{LZ_0}\left(1-2y\right)\ee
Therefore, to $O(h)$, the required coordinate transformation is
\B
\label{order h result}
Z(y,t)&=&Z_0+\frac{ht}{L}\left(1-2y\right)+O(h^2)\\ T(y,t)&=&t+O(h^2).\nn\E
The metric (\ref{yt metric}) is then given by
\B
\label{order h metric}
\d s^2&=&\left(4\frac{t^2 h^2}{Z_0^2}+O(h^3)\right)\d y^2 + O(h^2)\,\d y
\d t\\
&&\hspace{-3mm}+\left(Z_0^2+2\frac{Z_0
  t}{L}\left(1-2y\right)h+O(h^2)\right)\left(-\d t^2+\d\x^2\right)\nn\E
which, to $O(h)$, is of the required form (\ref{fixed gauge}). As a
check, we can calculate the proper times on the two branes,
\be
\label{proper times}
\d t_\pm=\left(Z_0\pm\frac{ht}{L}+O(h^2)\right)\d t\ee
giving the Hubble parameters as
\be
\label{hubble}
H_\pm=\pm\frac{h}{L Z_0^2}+O(h^2).\ee
The function $d(t)$, giving the proper distance between the branes
along lines of constant $y$, can be read off from (\ref{order h
  metric}) as
\be
\label{d(t) order h}
d(t)=\frac{2t}{Z_0}h+O(h^2)\ee
which gives
\be
\label{check}
H_\pm=\pm\frac{1}{2L}\frac{\d}{\d t_\pm}d(t)+O(h^2),\ee
in agreement with the low-energy result  of \cite{deRham:2005xv}.

\subsection{Higher orders}
The analysis at higher orders is straightforward. The constraint
$\partial_y g_{yy}=0$ reduces to an equation of the form
\be
\label{Z'' general n}
Z''_n(y,t)=\sum_{i,j}^{\mathrm{fin}} c^{(n)}_{ij} t^i y^j\ee
where the superscript means that the sum is finite, i.e. a binomial in
$y$ and $t$. This can easily be integrated to obtain $Z_n(y,t)$ up to two
arbitrary functions of $t$, which are fixed by the boundary conditions
(\ref{Z bc}) and the requirement that $Z(y,0)=Z_0$. Similarly, the
other constraint $g_{yt}=0$ is of the form
\be
\label{T' general n}
T'_n(y,t)=\sum_{i,j}^\mathrm{fin} d^{(n)}_{ij} t^i y^j\ee
giving $T_n(y,t)$ up to an arbitrary function of $t$. This is a
manifestation of the $t$-reparameterisation invariance, and can be
fixed for example by setting $T(t,0)=t$. In this particular gauge, the
next two sets of functions are
\B
\label{z2}
Z_2(y,t)&=&\frac{2t^2}{L^2Z_0^2}y(y-1)\\
\label{t2}
T_2(y,t)&=&-\frac{2t}{Z_0^4}y(1-y)\\
\label{z3}
Z_3(y,t)&=&\frac{t}{L^3 Z_0^5}\left\{4\left(L^2+t^2
  Z_0^2\right)\left(\frac{y^2}{2}-\frac{y^3}{3}\right)\right.\\
&&\qquad\quad\left. + y\left(\frac{7}{12}L^2-\frac{2}{3}t^2
  Z_0^2\right)-\frac{5}{8}L^2\right\}\nn\\
\label{t3}
T_3(y,t)&=&\frac{2t^2 y}{3LZ_0^5}\left(9-12y+8y^2\right).
\E
This transformation satisfies all the required boundary conditions,
as well as the conditions on the metric (\ref{fixed gauge}), up to $O(h^3)$:
\begin{align}
g_{yy}&=4\frac{t^2 h^2}{Z_0^2}+\frac{h^4
  t^2}{3L^2Z_0^6}\left(8\,t^2\,Z_0^2-19L^2 \right) +O(h^5)\nn\\
g_{yt}&=O(h^4)\nn
\end{align}
In order to prove the relations (\ref{Z'' general n}) and (\ref{T'
  general n}), we use induction. Specifically, we aim to prove that
  the general form of $Z_n$ and $T_n$ is a binomial function of $t$
  and $y$ of the form
\begin{align}
\label{Z general n}
Z_n(y,t)&=t\sum_{i,j=0}^\mathrm{fin} a^{(n)}_{ij} y^i t^j\quad\mathrm{for\
}n\geq 1\\
\label{T general n}
T_n(y,t)&=y\,t\sum_{i,j=0}^\mathrm{fin} b^{(n)}_{ij} y^i t^j\quad\mathrm{for\
}n\geq 2.
\end{align}
From (\ref{z1}) and (\ref{z2}-\ref{t3}), we
see that this is indeed the case for $n\leq 3$. We now assume that we
have found the $Z_m$ and $T_m$ for $m<n$, satisfying the required
boundary conditions. We then look at the constraint that $g_{yy}$ is
independent of $y$, which becomes
\be
\label{dygyy}
Z Z_y Z_{yy}=Z_y^3+\frac{1}{L^2}\left(n(Z)n'(Z) Z^3 Z_y
T_y^2+n(Z)Z^3T_y T_{yy}\right).\ee
We aim for an equation of the form (\ref{Z'' general n}) and,
examining the LHS of this equation, we see that the lowest order in
$h$ at which $Z''_n$ appears is at $O(h^{n+1})$,
\B
Z Z_y Z_{yy}&\supset& \left(Z_0^3\,
Z_1'Z_n''+t^2 B(y,t) \right)h^{n+1}\nn\\
\label{zn'' term}
&=&\left(-\frac{2t}{L}Z_0^2 Z_n''+t^2 B(y,t)\right)h^{n+1}\E
where the symbolic notation $B(y,t)$ represents a binomial function in $t$ and
$y$ depending only on the binomial functions $Z_m$ for $m<n$. The factor of
$t^2$ multiplying this function is displayed explicitly since we need
to keep track of powers of $t$ to ensure the RHS of (\ref{Z'' general
  n}) is indeed a binomial with no negative powers of $t$.
Now starting on the RHS of (\ref{dygyy}), we note that
\[
Z_y^3=Z_0^3\sum_{a,b,c=1}^\infty h^{a+b+c}Z'_a\,Z'_b\,Z'_c\]
and hence, at $O(h^{n+1})$, does not contain $Z'_n$, only the lower
$Z_m$. Hence we can write symbolically
\[
Z_y^3=t^3 B(y,t).\]
Both $T_y^2$ and $T_y T_{yy}$ have a factor $t^2\,h^4$ in
front, and can be written with the inductive hypothesis as
\be
\label{tt}
T_y^2,\ T_yT_{yy}\ \sim\  t^2 h^4 B(y,t)+ t^2 O(h^{n+2}).\ee
Furthermore, as in (\ref{function expansion}), a general analytic
function of $Z$ can be written
\be
\label{f(Z)}
f(Z)=f(Z_0)+\sum_{k=1}^\infty
h^k\left(\sum_{m=1}^k\sum_{i_j>0,\sum{i_j}=k}
Z_{i_1}...Z_{i_k}\right)\ee
Combining (\ref{tt}) and (\ref{f(Z)}), we see that the term with
coefficient $1/L^2$ on the RHS of (\ref{dygyy}) contributes, at
$O(h^{n+1})$, a term of the form $t^2 B(y,t)$.

Hence, putting all of the above together, we have shown that $Z_n$ is given in terms of the $Z_m$ for
$m<n$ by an equation of the form (\ref{Z'' general n}) with an
additional pre-factor of $t$. This equation
can always be integrated to give
\be
\label{expression for Z_n}
Z_n(y,t)=t B_n(y,t)+a_n(t) y+b_n(t)\ee
where $B_n(y,t)$ is a binomial expression depending on the $Z_m$ for
$m<n$, and $a_n(t)$ and $b_n(t)$ are arbitrary functions, which are
then uniquely fixed by the boundary conditions. Unlike the expansion
in $t$, we can in principle calculate all the $Z_n$ analytically,
though this is not a terribly helpful procedure. What is important is
that the boundary conditions are trivial to implement - (\ref{Z bc}) will
prescribe values for $Z'_n$ at $y=0$ and $y=1$, which, together with
$Z(y,0)=Z_0$, fix $a_n(t)$ and $b_n(t)$ uniquely from (\ref{expression for Z_n}).

To show the existence of the $T_n$, we need to examine the equation
\be
\label{t equation}
T_y\,T_t=\frac{L^2}{Z^2 n(Z)^2}Z_y Z_t\ee
which follows from the metric (\ref{yt metric}). $T_n$ enters the LHS
at $O(h^n)$, 
\[
T_y\,T_t\supset h^n\left(T'_n+B(y,t)\right)\]
where again $B(y,t)$ stands for binomial terms composed of
already-known functions, the $T_m$ for $m<n$ in this case. In fact the
RHS of (\ref{t equation}) does not contain $Z_n$ at $O(h^n)$, since
both $Z_y$ and $Z_t$ are $O(h)$. Therefore, the equation for $T_n$ is
of the form (\ref{T' general n}), specifically
\[
T'_n(y,t)=t\,C_n(y,t)\]
with the factor of $t$ coming from the $Z_y$. This can be integrated to
\be
\label{expression for T_n}
T_n(y,t)=y\,t\, \tilde{C}_n(y,t)+d_n(t)\ee
where once again $\tilde{C}_n$ is a binomial function of $y$ and $t$. The
arbitrary function $d_n(t)$ encodes the $t$ reparameterisation invariance
of the gauge. Because of the pre-factor $y$ in (\ref{expression for
  T_n}), the gauge $T(0,t)=t$ corresponds to setting $d_n(t)=0$.
We have therefore shown that there exist simple, binomial expressions
for $T_n(y,t)$ and $Z_n(y,t)$ for all $n$, with an extremely simple if
tedious procedure for finding them explicitly. 

It should be emphasised, as in \S\ref{t expansion}, that the precise
nature of the boundary conditions are not important. Hence, although
the explicit solution constructed in this section corresponds to
empty, fine-tuned branes, it would be simple in principle to repeat
the process for more general situations with both de-tuned branes and
matter at the background level.

\section{Cosmological Perturbations}
\label{perturbations}
We now consider more general spacetimes, where the spatial homogeneity
is broken by cosmological perturbations. Our starting point is to
consider a general background spacetime, taking the metric to be of the form
\be
\label{gn background}
\d s_0^2=\d Y^2-n(Y,T)^2\d T^2+a(Y,T)^2 \Omega_{ij}\d x^i \d x^j\ee
where $\Omega_{ij}$ is the relevant maximally-symmetric spatial
metric of constant curvature. In this frame the branes are taken to
have loci $Y_\pm(T)$. We then assume that a coordinate transformation
\be
\label{Y and T}
T=f(y,t),\quad Y=g(y,t)\ee
can
be constructed as described previously, satisfying
\be
\label{cond 1}
g_y g_t-n(f,g)^2 f_y f_t = 0,\ee
\be
\label{cond 2}
g_y^2 - n(f,g)^2 f_y^2 = d(t)^2,\ee
with boundary conditions
\[
g\left(y=0,t\right)=Y_-\left(f(y=0,t)\right),\]
\[ g\left(y=1,t\right)=Y_+\left(f(y=1,t)\right),\]
bringing the metric into the standard form
\be
\label{fixed background}
\d s_0^2=d(t)^2 \d y^2 - b(y,t)^2 \d t^2 + a(y,t)^2 \Omega_{ij} \d x^i \d x^j\ee
with the branes fixed at $y=0$ and $y=1$ (the precise forms of
$b(y,t)$ and $a(y,t)$ are not important). Note that, in this section,
we shall not fix the $t$-reparameterisation invariance as in (\ref{t
  gauge fix}), which will allow us to consider more general situations
than the branes moving apart after a collision, such as perturbations
about static branes.

We now allow the presence of cosmological perturbations, both to the
background metric (\ref{gn background}) and, in the case of scalar
perturbations, to the positions of the branes themselves. We then
attempt to find an infinitesimal gauge transformation which will not
only retain the properties
\be
\label{requirement}
g_{5\mu}=\frac{\partial}{\partial y}g_{55}=0\ee
of the metric, but bring the brane positions back to $y=0$ and $y=1$
if necessary. Firstly, we
shall perform the coordinate transformation (\ref{Y and T}) to bring
the background into the form (\ref{fixed background}), before
constructing the gauge transformation required to accommodate both the
brane positions and the bulk metric perturbations into the SFB
gauge at first order. Note that, in this section, we are still not considering any
non-gravitational degrees of the freedom in the bulk, which would in
principle also need perturbing, such as a bulk scalar field.

\subsection{Tensor perturbations}
Firstly, we consider the simplest case of tensor perturbations. In
five dimensions, the most general infinitesimal tensor perturbation
\cite{vandeBruck:2000ju} to (\ref{gn
  background}) only affects the transverse spatial components, i.e.
\be
\label{tensor}
g_{ij}\rightarrow a(Y,T)^2\left(\Omega_{ij}+h_{ij}\right)\d x^i \d x^
j.\ee
This will therefore not affect the properties (\ref{requirement}) of
the metric in the new frame (\ref{fixed background}). Furthermore, the
brane trajectories are scalars, and are therefore not affected by a
purely tensor perturbation. Hence tensor perturbations can trivially
be accommodated into the gauge (\ref{fixed background}). 

\subsection{Vector perturbations}
There are six vector degrees of freedom for five-dimensional
perturbations about the background (\ref{gn background}), encoded in three divergenceless vectors:
\B
\label{vector}
\d s^2&=&\d s_0^2 + Q^{(T)}_i \d x^i \d T + Q^{(Y)}_i \d x^i  \d Y\\
&&+a(Y,T)^2 F_{(i|j)}\d x^i \d x^j\nn\E
where $|$ denotes the covariant derivative with respect to the metric
$\Omega_{ij}$ and $F^i_{|i}=Q^{(Y)i}_{|i}=Q^{(T)i}_{|i}=0$. Performing
the background coordinate transformation (\ref{Y and T}), we obtain
the perturbed version of (\ref{fixed background})
\B
\d s^2&=& d(t)^2 \d y^2 -b(y,t)^2\d t^2 +\left(Q^{(T)}_i f_y + Q^{(Y)}_i g_y\right) \d y \d x^i\nn\\
&&+\left(Q^{(T)}_i f_t + Q^{(Y)}_i g_t\right) \d t \d x^i\nn\\&&+a(y,t)^2\left(\Omega_{ij}+2 F_{(i|j)}\right)\d x^i \d x^j\nn\E
Therefore, in order to satisfy the gauge conditions (\ref{requirement}), we require the linear combination
\be
\label{vector combination}
Q^{(T)}_i f_y + Q^{(Y)}_i g_y\ee
to vanish. Under the most general infinitesimal vector gauge transformation
\[
x^i\rightarrow x^i + \eta^i,\quad \eta^i_{|i}=0,\]
the three vector perturbations transform as
\B
\delta Q^{(Y)}_i&=&-a(y,t)^2 \partial_y \eta_i\nn\\
\delta Q^{(T)}_i&=&-a(y,t)^2 \partial_t \eta_i\nn\\
\delta F_i&=&-\eta_i\nn\E
where $\eta_i=\Omega_{ij}\eta^j$. There are therefore two gauge-invariant vector perturbations (i.e. four degrees of freedom) corresponding to $Q^{(Y)}_i-a^2 F'_i$ and $Q^{(T)}_i-a^2 \dot{F}_i$, and any one vector perturbation can be set to zero. In particular, we can remove the linear combination (\ref{vector combination}) with suitable $\eta_i$ satisfying
\[
f_y \partial_y \eta_i+g_y \partial_t \eta _i=\frac{1}{a(y,t)^2} \left(Q^{(T)}_i f_y+Q^{(Y)}_i  g_y\right).\]
Again there is no vector perturbation to the brane trajectories, hence these still lie at $y=0$ and $y=1$.

\subsection{Scalar perturbations}
Scalar perturbations require a more detailed treatment, both because there are more of them in general and because one must also assume that the brane positions are perturbed. In five dimensions there are seven scalar degrees of freedom,
\B
\label{scalar}
\d s^2&=&\d s_0^2+\phi_{55}\d Y^2+2\phi_{05}\d Y \d T-n^2(Y,T)\phi_{00}\d T^2\nn\\
&&+Q^{(Y)}_{,i} \d Y \d x^i + Q^{(T)}_{,i}\d T \d x^i\nn\\
&&+2\,a(y,t)^2\left(2\,\psi+E_{,ij}\right)\d x^i \d x^j\E
Three of these can be removed at this stage by a scalar gauge transformation; effectively, this is just the statement that one can always choose a Gaussian-Normal gauge for the bulk where the off-diagonal  terms $Q^{(Y)}_{,i}$, $\phi_{05}$ and the $g_{55}$ perturbation vanish:
\B
\label{gn scalar}
\d s_{GN}^2&=&\d s_0^2-n^2(Y,T)\phi\,\d T^2+ Q^{(T)}_{,i}\d T \d x^i\nn\\&&+2\,a(y,t)^2\left(2\,\psi+E_{,ij}\right)\d x^i \d x^j\E
where $\phi\equiv\phi_{00}$. Performing the coordinate transformation (\ref{Y and T}), we find
\B
\d s^2&=&\left[d(t)^2-n^2 f_y^2\phi\right]\d y^2-2n^2 f_y f_t\phi\, \d y \d t + Q_{,i} f_y \d x^i \d y\nn\\
&&-\left(b^2+n^2 f_t^2\phi\right)\d t^2+Q_{,i}f_t \d x^i \d t\nn\\
\label{fixed scalar metric 1}
&&+a^2\left[(1+\psi)\Omega_{ij}+E_{,ij}\right]\d x^i \d x^j\E
where all the problematic terms have been written out in the first line, and $n=n(y,t)\equiv n\left(f(y,t),g(y,t)\right)$ etc. The background brane positions are fixed in this frame, but in the presence of scalar perturbations will be displaced to
\be
\label{brane positions}
y_{+}\left(\x,t\right)=\delta y_{+}\left(\x,t\right),\quad y_{-}\left(\x,t\right)=1+\delta y_{-}\left(\x,t\right)\ee
In five dimensions there are four gauge-invariant scalar degrees of
freedom, hence we cannot eliminate any more of these perturbation
variables with gauge transformations. We can, however, perform another
gauge transformation to rearrange them in such a way as to satisfy the
conditions (\ref{requirement}), whilst simultaneously bringing the two
branes back to the fixed positions $y=0$ and $y=1$. Under the most
general infinitesimal scalar coordinate transformation
\be
\label{scalar transformation}
x^a\rightarrow x^a+\xi^a,\quad \xi^a=\left(\xi,\chi,\partial^i\zeta\right),\ee
the metric perturbations transform as
\B
\delta g_{55}&=&-2 d^2 \xi'-2d\dot{d}\,\chi\nn\\
\delta g_{50}&=&-d^2 \dot{\xi}+b^2 \chi'\nn\\
\delta g_{5i}&=&-\partial_i \left[\zeta'+d^2\xi-2\frac{a'}{a}\zeta\right]\nn\E
where $'=\partial_y$ and $\dot{}=\partial_t$. Hence, from (\ref{fixed scalar metric 1}), we find the three conditions necessary to fix the required gauge:
\B
\label{xi}
&&2d^2\xi''+2d\dot{d}\,\chi'+\left(n^2\,f_y^2\,\phi\right)'=0,\\
\label{zeta}
&&\zeta'+d^2\xi-2\frac{a'}{a}\zeta-Q f_y=0\\
\label{chi}
&&b^2\chi'-d^2\dot{\xi}-n^2\,f_y\,f_t\,\phi=0,\E
together with the boundary conditions
\be
\label{xi bc}
\xi\left(y=0\right)=-\delta y_+\left(\x,t\right),\quad
\xi\left(y=1\right)=-\delta y_-\left(\x,t\right)\ee
which, from (\ref{brane positions}), will bring the brane positions to
$y=0$ and $y=1$. $\zeta$ and $\chi$ have no boundary conditions to satisfy; once one has found the required $\xi$, they follow immediately from (\ref{zeta}) and (\ref{chi}) respectively. Eliminating $\chi'$ from (\ref{xi}) with (\ref{xi}), we obtain the equation of motion for $\xi$,
\begin{align}
\label{xi eqn}
\xi''+k(y,t)\dot{\xi}&=F(y,t,\x)\\
F(y,t,\x)&\equiv\frac{\left(n^2\,
  f_y^2\, \phi\right)'}{2d^2}-\frac{n^2d\dot{d}}{b^2\,t}\, f_y
f_t\,\phi\nn\\
k(y,t)&\equiv\frac{d\dot{d}}{b^2}\nn
\end{align}
Interestingly only one of the original seven possible five-dimensional
perturbations appears in this equation. In principle some information
about the values of $\phi$ and its derivatives at the boundaries can
be inferred from the Isra\"el junction conditions, though in practice
  this is not relevant. The perturbation is absorbed into a `forcing'
  term $F(y,t,\x)$, with (\ref{xi eqn}) taking the form of a
  generalised diffusion equation.

\subsubsection{Perturbations about static branes}
For static Randall-Sundrum branes (e.g. $C=0$ for empty branes), the
original background configuration (\ref{gn background}) is already in an SFB gauge with $d(t)=1$. In this simple case, the equation of motion (\ref{xi eqn}) reduces to
\be
\label{static xi equation}
\xi''=F(y,t,\x)\ee
The general solution of this equation is
\be
\label{static xi solution}
\xi(y,t,\x)=\xi^\mathrm{P}(y,t,\x)+(1-y)A(\x,t) +y\,B(\x,t),\ee
where $\xi^\mathrm{P}$ is a particular solution to (\ref{static xi
  equation}), and the linear term represents the most general
homogeneous solution. $A(\x,t)$ and $B(\x,t)$ are arbitrary functions
which can be set so that $\xi$ satisfies arbitrary boundary conditions
at $y=0$ and $y=1$. In particular, one can always solve the boundary
conditions (\ref{xi bc}) with the choice
\B
\label{A and B}
A(\x,t)&=&-\left(\delta y_+(\x,t)+\xi^\mathrm{P}(0,t,\x)\right)\\
B(\x,t)&=&-\left(\delta y_-(\x,t)+\xi^\mathrm{P}(1,t,\x)\right)\nn\E
Hence any cosmological perturbations about a static, fixed-brane
background can always be accommodated into an SFB gauge.

\subsubsection{Perturbations about general backgrounds}
For non-trivial $k(y,t)$ the process of constructing a solution to
(\ref{xi eqn}) for arbitrary boundary conditions (\ref{xi bc}) is far
more complicated and one must, once again, resort to a series solution.
The boundary conditions (\ref{xi bc}) can be dealt with by a simple
redefinition of the forcing function $F(y,t,\x)$, by setting
\be
\label{remove bc}
\xi(y,t,\x)=w(y,t,\x)-(1-y)\delta_+(\x,t)-y\,\delta_-(\x,t).\ee
The new function $w$ has the simpler boundary conditions that $w=0$ at
$y=0$ and $y=1$, and satisfies the equation
\be
\label{new equation}
w''+k(y,t)\dot{w}=G(y,t,\x)\ee
where
\[
G(y,t,\x)\equiv F(y,t,\x)+k(y,t)\left((1-y)\delta y_+(\x,t)+y\,\delta y_-(\x,t)\right)\]
We now expand the functions $G$, $w$ and $k$ in powers of $t$. Considering the case once again where the branes are moving apart
after a collision which occurred at $t=0$, we see from its definition
in (\ref{xi eqn}) that $k(y,0)=0$, hence there is no $t^0$ term in its
expansion. We therefore write
\B
k(y,t)&=&\sum_{t=1}^\infty t^n k_n(y)\nn\\
G(y,t,\x)&=&\sum_{t=0}^\infty t^n G_n(y,\x)\nn\\
w(y,t,\x)&=&\sum_{t=0}^\infty t^n w_n(y,\x),\nn\E
which, upon substitution into (\ref{new equation}), gives
\B
\label{w0}
w''_0&=&G_0\\
\label{wn}
w''_n+n k_1 w_n&=&G_n-\sum_{m=1}^{n-1} m\, w_m\, k_{n+1-m}\equiv
\tilde{G}_n\E
together with the boundary conditions
\[w_n(0,\x)=w_n(1,\x)=0.\]
(\ref{wn}) is just a simple, linear Sturm-Liouville problem which can
always be solved with an eigenfunction expansion subject to certain
conditions. Specifically, we consider the eigenfunction equation
\be
\label{eigenvalue equation}
\phi_m''(y)+\lambda_m k_1(y)\phi_m(y)=0,\quad\phi_m(0)=\phi_m(1)=0.\ee
The eigenfunctions $\phi_m$ for $m=1,2,3...$ will form a complete,
orthogonal basis, and so we can expand
\B
\tilde{G}_n(y,t,\x)&=&k_1(y)\sum_m G_{nm}(\x,t)\,\phi_m(y),\nn\\
\quad w_n(y,t,\x)&=&\sum_m w_{nm}(\x,t)\,\phi_m(y)\nn\E
and these sums converge absolutely. Substituting these expressions
back into (\ref{wn}), we find
\be
\label{coefficients}
w_{nm}=\frac{G_{nm}}{n-\lambda_m}\ee
and hence the required solution to (\ref{xi eqn}) subject to the
boundary conditions (\ref{xi bc}) has been constructed. From
(\ref{coefficients}), we see that this solution only exists if the
eigenvalues of (\ref{eigenvalue equation}) are either never
integer-valued, or, if $\lambda_m=N$ for some integer $N$, then the
corresponding $G_{Nm}$ vanishes.

Note that this conclusion is consistent with
$t$-reparameterisation. For example, if we reparameterise
\[t\rightarrow T(t)=T_1\, t+T_2\, t^2+...\]
then 
\[
k_1(y)\rightarrow \frac{k_1(y)}{N}\]
where $N$ is the minimal positive integer such that $T_N\neq 0$. The
eigenvalues of (\ref{eigenvalue equation}) will therefore all be
multiplied by $N$, and $G_{nm}$ is only non-zero for $n$ a multiple of
$N$. Hence the conclusion remains the same. This condition on the functions $k(y,t)$ and $G(y,t,\x)$ can be regarded
as necessary and sufficient for scalar perturbations to be accommodated within the
SFB gauge. 

Note that if $k_1(y)=0$, the above argument will simplify considerably
since each equation for $w_n$ will just be of the form
\be
w''_n=\tilde{G}_n\qquad\mathrm{for\ }k_1=0\nn\ee
which can be integrated immediately, with two arbitrary constants to
fix the boundary conditions. Hence, in this case, it will always be
possible to find the required gauge. However, for branes moving apart
after a collision, from the above argument $k_1(y)$ will be
proportional always to its value $1/b(y,0)$ in the gauge $d(t)=t$, which
will not vanish. Hence one will have to check the eigenspectrum of the
problem to verify the existence of the required SFB gauge.

The precise nature of the function $G(y,t,x)$ encoding both the bulk
perturbation $\phi$ and the perturbations $\delta y_\pm$ to the brane
positions is not important for the above discussion. Hence one would
expect, subject to the above caveats, to be able to include the
effects of matter perturbations on the brane if desired. Since the
perturbative construction of the background coordinate transformation
in \S\S\ref{t expansion} and \ref{C expansion} were not sensitive to
the precise nature of the boundary conditions either, one can expect
to include matter at the background level as well. Hence we have
presented in this paper a method for constructing SFB gauges very generally for Randall-Sundrum cosmologies.

\section{Conclusions}
The aim of this paper was to demonstrate the existence of gauges for
Randall-Sundrum two-brane braneworlds in which the metric takes the
particularly simple form (\ref{fixed gauge}) whilst, at the same time,
the branes are located at fixed positions. Such gauges have been
assumed to exist already in the literature
\cite{Kanno:2002ia2,deRham:2005qg,deRham:2005xv,Shiromizu:2002ve},
without proof, since they simplify considerably the analysis of the
five-dimensional Einstein equations necessary to derive analytic
results for effective theories. 

We first examined the dynamics of the system in the background,
deriving from the trajectories of the branes the necessary boundary
conditions for the coordinate transformation functions. These were
then found using two different power series expansions, expanding
either in time or in the parameter $\sqrt{C}$ related to the mass of
the bulk black hole.

The former approach corresponds to an expansion in the radion, the
proper distance between the two branes, and is adapted ideally to the
case of two branes moving apart soon after a collision. This regime is
of interest in recent attempts to model the Big Bang as a collision
between two branes
\cite{Khoury:2001wf,Khoury:2001bz,Gibbons:2005rt,Jones:2002cv,Turok:2004gb,Tolley:2003nx,Steinhardt:2002ih,Kanno:2002py}.
However, the non-linear equations which one needs to solve order by
order in $t$ do not admit an analytic solution. One can, however,
demonstrate the existence of all terms in the series, satisfying the
required boundary conditions, corresponding to being able to perform
the necessary coordinate transformation correct to $O(t^N)$ for
arbitrary $N$. Whilst no analytic information about the convergence of
the series is possible, numerical simulations show good agreement with
the exact results, and we conjectured that the radius of convergence
of the series would be equal to $t_h$, the time taken for the first
brane to hit the horizon.

A second method, namely an expansion in $\sqrt{C}$, corresponds to an
expansion in the velocity of the branes. It is possible to obtain
analytic expressions for the coordinate transformation functions, and
the resulting SFB metric, correct to arbitrary order in $h$. This is
because the equations at each order are simple, linear differential
equations, involving just binomial functions which are trivial to integrate.

Results on cosmological backgrounds are of limited interest; the
important question is whether or not one can find an infinitesimal
gauge transformation to accommodate arbitrary cosmological
perturbations, at first order, into an SFB gauge. We addressed this
question last, and showed that this is trivially the case for tensor
and vector perturbations. Scalar perturbations require considerably
more work because one must also take into account a perturbation of
the brane trajectories themselves, and hence impose boundary
conditions on the infinitesimal coordinate transformation $\xi=\delta
x^5$. The second-order partial differential equation for $\xi$ can be
solved by an eigenvalue expansion, giving a well-defined constraint on the eigenvalues of (\ref{eigenvalue equation}) for the SFB
gauge to exist.

Whilst the arguments in this paper do not apply to braneworlds with a
`non-empty' bulk (i.e. with non-gravitational bulk degrees of freedom), they are not sensitive to the precise nature of the
Israel junction conditions; one can include both detuned tensions and
brane matter without altering the conclusions.

\begin{center}
\textbf{Acknowledgements:}
\end{center}
The author would like to thank Anne Davis for her supervision and
Claudia de Rham for her considerable help in the preparation of this
manuscript. He is supported by Trinity College, Cambridge.

\end{document}